# CURRENT INJECTION ATTACK AGAINST THE KLJN SECURE KEY EXCHANGE

**Hsien-Pu Chen[1), Muneer Mohammad[2)], and Laszlo B. Kish[2)]**

1) Texas A&M University, Department of Electrical and Computer Engineering, College Station, TX 77843-3128, USA,
  (✉ barrychen@tamu.edu, +1-979-633-4850)
2) Texas A&M University, Department of Electrical and Computer Engineering, College Station, TX 77843-3128, USA
  (muneer@tamu.edu, Laszlo.Kish@ece.tamu.edu)

**Abstract**

The Kirchhoff-law-Johnson-noise (KLJN) scheme is a statistical/physical secure key exchange system based on the laws of classical statistical physics to provide unconditional security. We used the LTSPICE industrial cable and circuit simulator to emulate one of the major active (invasive) attacks, the current injection attack, against the ideal and a practical KLJN system, respectively. We show that two security enhancement techniques, namely, the instantaneous voltage/current comparison method, and a simple privacy amplification scheme, independently and effectively eliminate the information leak and successfully preserve the system's unconditional security.

Keywords: KLJN, current injection attack, secure key exchange, unconditional security, privacy amplification.

## 1. Introduction

Unconditional security means that, even in the case of a perfectly able eavesdropper (Eve), the perfect security limit (zero information for Eve) of communication can be approached if sufficient resources (time, etc.) are available [1]. Unconditional security is essential in intelligent vehicle systems [2,3]; for power and sensor networks of strategical importance [4,5]; for ultra-strong PUF hardware keys [6]; and in secure computer, instrument and video game systems [7].

Currently, the only unconditionally secure key exchange that can be integrated on a chip and has reasonable price is the Kirchhoff-law-Johnson-(like)-noise (KLJN) scheme, which was first introduced in 2005 [8-11]. It is the only classical physical competitor of quantum communicators [1]. Its security is based on the Fluctuation-Dissipation Theorem [9] of classical statistical physics and the properties of Gaussian stochastic processes [12]. There have been various valid attacks causing minor information leak but not a full crack, such as methods using the cable capacitance [13], cable resistance (Bergou-Scheuer-Yariv attack) [14-18], temperature-inaccuracy (Hao-attack) [19-21]. However, in each case, the information leak can be eliminated whenever sufficient resources (either specific hardware, higher accuracy, or enough time for privacy amplification) are available, thus the system stays unconditionally secure [1].

Some other attacks are simply invalid with fundamental flaws in their model and physics. Yet the analysis of these faulty attempts in the subsequent rebuttals [23-25,28,30] provides deeper understanding of the security of the KLJN scheme. Perhaps the best example is the Gunn-Allison-Abbott (GAA) "directional coupler" attack published in one of the Nature journals [22], where serious conceptual and theoretical errors [23-25] incorrectly imply that a directional coupler can be built and that will serve with information leak. However, directional coupler cannot be built for the KLJN's no-wave (quasi static) situation [28], moreover, even an existing directional coupler would be insufficient to extract any information in the steady state [23-25]. Interestingly, perhaps as the result of wishful thinking, the mistakes mentioned above were "verified" by experiments with severe flaws in [22], where even the KLJN loop was broken into two coupled Kirchhoff loops by a shunt resistor a the middle, see the rebuttal in [26]. Two coupled Kirchhoff loops have never been claimed secure and they obviously represent a giant information leak. Actually, any deviation from the original single KLJN loop implies information leak, which sometimes can be eliminated by introducing proper modifications, see [18].

Another one, a high-profile, many-sided cracking attempt by Bennett-Riedel [27], while it triggered useful and extensive analysis in a rebuttal [28], it has also failed with all of its goals, further indicating that physical security is a subtle topic. Here we also mention an earlier unsuccessful attempt [29], which, similarly to the above ones, triggered discussions [30] with valuable outcomes. Finally, we acknowledge a recent transient attack by GAA [31], which is valid, even though there are severe mistakes [44] in the considerations about security and physics in the Appendix of the paper, and a simple, known solution [32] does exist to fully eliminate this attack.

In conclusion, the unconditional security of the KLJN scheme remains unchallenged. As with the evolution of quantum communicators, further attacks schemes are expected to emerge and to trigger new defense solutions that eliminate those attacks, too.

The core KLJN secure key exchange system [1,9-11,32-40] is shown in Fig. 1, while [2-7] and [41-43] are dealing with advanced aspects with expansions and applications. At the beginning of each bit exchange period (BEP), Alice and Bob randomly select a resistor from the set $\{R_L, R_H\}$, $R_L \neq R_H$, where $R_L$ represents the Low bit value (L) and $R_H$ the High bit value (H), and they connect the chosen resistors to the wire channel (cable). The Gaussian voltage noise generators emulates the Johnson noise of the resistors and deliver band-limited white noise with publicly agreed bandwidth and temperature $T_{\text{eff}}$. Within each BEP, Alice and Bob measure the current and voltage noises, $I_{\text{ch}}(t)$ and $U_{\text{ch}}(t)$, in the cable. Using the Johnson formula, they derive the unknown resistance value at the other end of the cable which is the difference between their own resistance and the total loop resistance [9]. Though Eve can also obtain the total loop resistance, she cannot distinguish the LH and HL bit situations, which indicates a secure bit exchange. The HH and LL bit situations are disregarded.

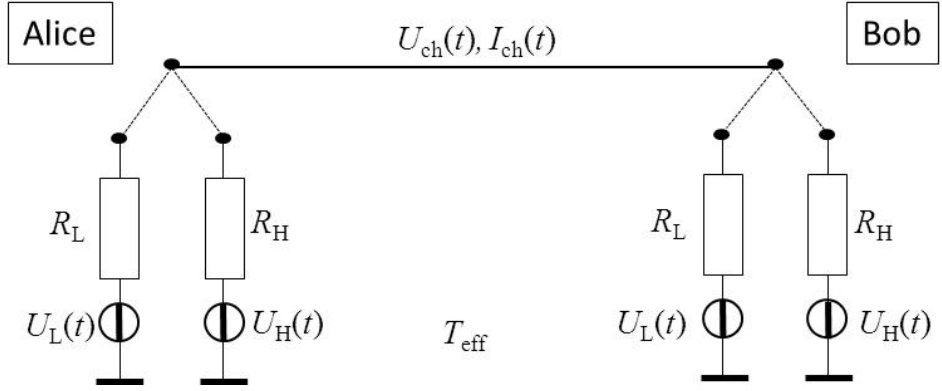

Fig. 1. Schematics of the Kirchhoff-law-Johnson-(like)-noise (KLJN) secure key exchange system. The resistor values are $R_L$ and $R_H$. The thermal noise voltages, $U_L(t)$ and $U_H(t)$, are generated at an effective temperature $T_{eff}$. The channel noise voltage and current are $U_{ch}(t)$ and $I_{ch}(t)$, respectively.

## 2. Current Injection Attack

The current injection attack is an active (invasive) attack, which was introduced in 2006 [9]. Its security analysis was given in 2013 [28] but the attack itself had never been practically tested.

### 2.1. The Attack Protocol

For the sake of simplicity but without losing generality, fixed LH bit arrangement with $R_L < R_H$ is assumed. During the exchange of the bit, Eve attempts to identify the location of $R_L$ and $R_H$ by injecting a Gaussian current $I_{inj}(t)$ of the same bandwidth as the channel noises into the cable while she measures the following cross-correlations during the exchange of the $i$-th key bit:

$$\rho_i^a = \langle I_{inj}(t) I_{cha}(t) \rangle_\tau, \tag{1}$$

$$\rho_i^b = \langle I_{inj}(t) I_{chb}(t) \rangle_\tau, \tag{2}$$

where $I_{cha}(t)$ and $I_{chb}(t)$ are the channel currents at Alice's and Bob's ends, respectively, see Fig. 2. The time average $\langle \ \rangle_\tau$ is taken over the bit exchange period $\tau$. According to the current divider rule, a greater current flows to the direction of the lower resistance. With Alice connecting to $R_L$ and Bob connecting to $R_H$, the cross-correlation $\rho_i^a$ at Alice's side is greater than the cross-correlation $\rho_i^b$ at Bob's side. For $N$ bits, Eve calculates $\rho_i = \rho_i^a - \rho_i^b$ ($i = 1,...,N$) and decides as follows:

If $\rho_i > 0$  then LH (*Eve guessed the bit correctly*), set $q_i = 1$ . (3)

If $\rho_i < 0$  then HL (*Eve guessed the bit incorrectly*), set $q_i = 0$ . (4)

When $N$ approaches infinity, the probability $p_E$ of Eve's successful guessing of the bits converges to the expected value of $q$ and

$$\langle q_i \rangle_N = p_E \text{ where } 0.5 \leq p_E \leq 1.$$ (5)

The case $p_E = 0.5$, indicates perfect security, that is, Eve's information is zero (equivalent to guessing the key bits by tossing an unbiased random coin [43]).

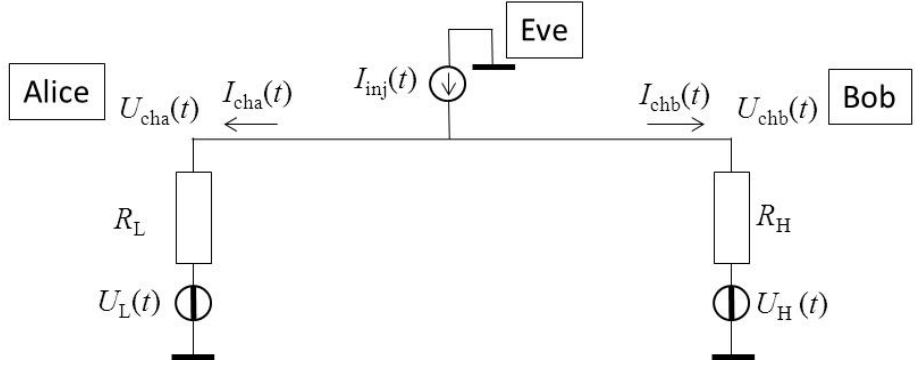

Fig. 2. Current injection attack against the ideal KLJN system [2]. $I_{inj}(t)$ is injection current. $I_{cha}(t)$, $I_{chb}(t)$, $U_{cha}(t)$ and $U_{chb}(t)$ are the channel currents/voltages at Alice's and Bob's ends respectively. (Note, the positive current directions at the two ends are chosen to follow the directions of the components of Eve's injected positive current).

## 2.2. Generic Defense Protocol

To provide security against the current injection attack, Alice and Bob can act similarly as against any active (invasive) attacks by measuring the instantaneous voltage and current amplitudes at their ends and compare them via public authenticated data exchange [1,10], see Fig. 3. In the case of deviance, Alice and Bob discard the bit or use a more advanced security protocol [1].

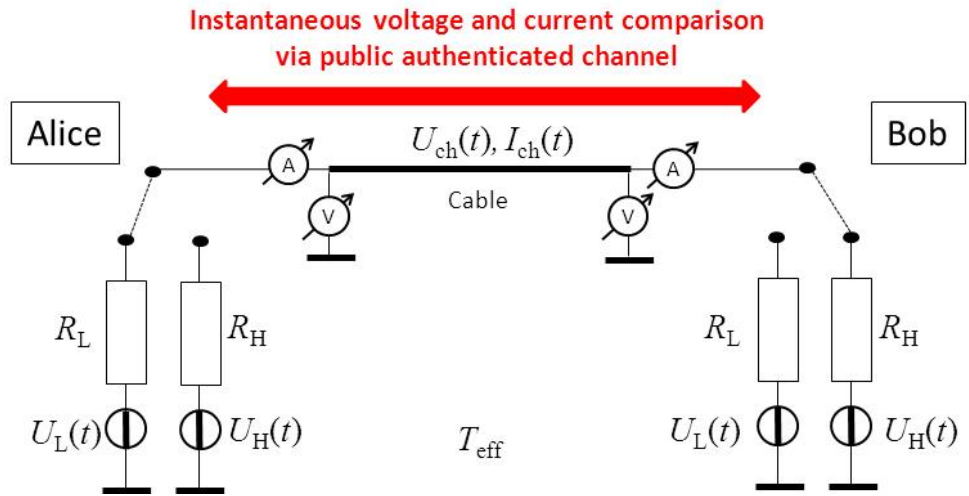

Fig. 3. The defense against the current injection attack.

## 3. Simulation Results

We used the RG58 coaxial cable model from the library of the cable and circuit simulator LTSPICE (Linear Technology), to test both the ideal and a practical KLJN system. We assumed that Alice and Bob selected $R_\text{L}=1\,\text{k}\Omega$ and $R_\text{H}=9\,\text{k}\Omega$, respectively; the bit exchange period $\tau$ was 0.1 s; $N$=10000; $T_\text{eff}=7.25\cdot10^{16}\,\text{K}$; and the bandwidth of the Gaussian noises 250 Hz.

We tested three levels of the injected Gaussian current noise, i.e., 0.1%, 1% and 10% of the rms channel current, in four different versions of the KLJN system (see Fig. 4). At each scenario, Eve's probability of guessing the bits was calculated, see Table 1.

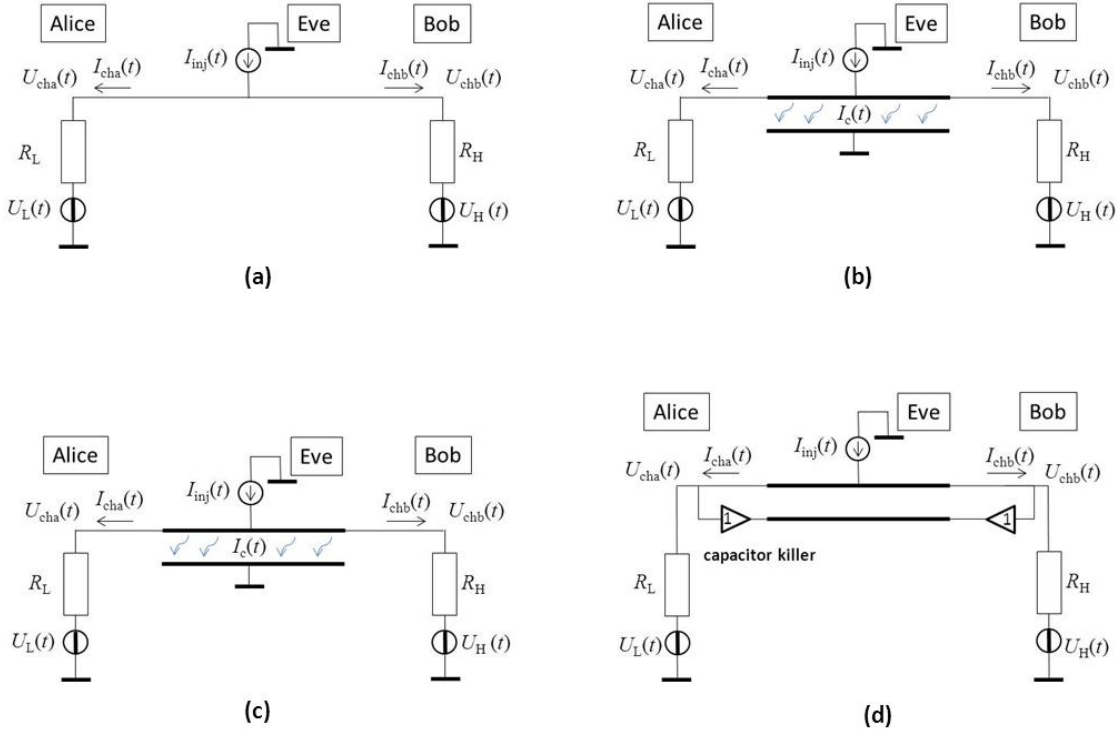

Fig. 4. The four different versions of KLJN system under the current injection attack. (a) the ideal KLJN system; (b) the practical KLJN system with 100 m cable; (c) the practical KLJN system with 1000 m cable; (d) the practical KLJN system with 1000 m cable and capacitor killer (ideal unity-gain voltage buffer) [13]. $I_c$ is capacitive current from the inner conductor to the outer shield of the cable. The cable is RG58 coaxial cable.

At 0.1% injected current level, in the ideal KLJN system, $p_E$ was 0.503, which is near to ideal. At 1% and 10% the information leak progressively increased with higher $p_E$ values (0.513 and 0.613). Eve's success probability values in the practical cable-based systems were very similar, see Table 1. Injecting even higher levels of current is also possible but that makes the detection of eavesdropping easier.

Table 1. Eve's success probability $p_E$ with 10000 bits key length.

| Injection current (in % of the rms channel current) | 0.1% | 1% | 10% |
| --- | --- | --- | --- |
| Ideal cable | 0.503 | 0.513 | 0.613 |
| 100 meters cable | 0.503 | 0.513 | 0.613 |
| 1000 meters cable | 0.501 | 0.510 | 0.608 |
| 1000 meters cable with capacitor killer | 0.503 | 0.513 | 0.613 |

## 4. Simulation Result of the Defense Methods

### *4.1. The defense protocols*

As mentioned above, in the ideal KLJN system, Alice and Bob can easily discover the current injection attack by comparing the instantaneous current data [9]. If the currents are different, Alice and Bob can discard the bit.

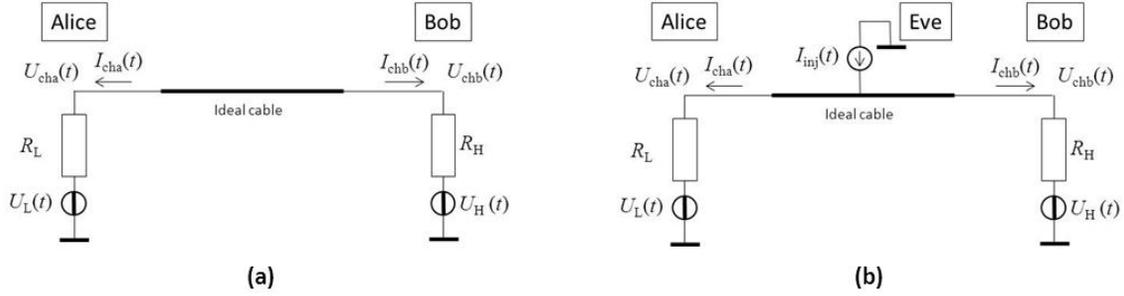

Fig. 5. Instantaneous voltage and current comparison against current injection attack in the ideal KLJN system. (a) No attack. (b) Under current injection attack.

However, in practical systems, the currents are slightly different due to the cable's capacitive current leak. Then Alice and Bob must also monitor and exchange the instantaneous voltage data, too. Then, they input the voltage data into the accurate cable model and compare the simulated currents $I^*_{cha}(t)$ and $I^*_{chb}(t)$ with the corresponding measured currents $I_{cha}(t)$ and $I_{chb}(t)$, see Fig. 6.

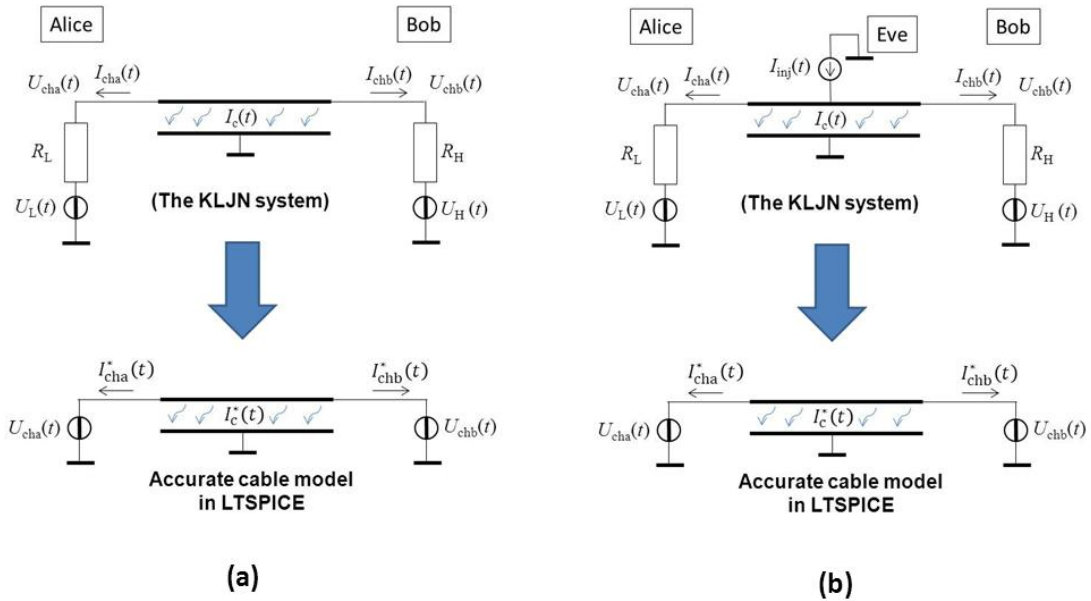

Fig. 6. The instantaneous voltage and current comparison against current injection attack in practical KLJN system: (a) No current injection attack, (b) Under current injection attack. $I^*_{cha}(t)$ and $I^*_{chb}(t)$ are the simulated currents at Alice's and Bob's side respectively. $I_c(t)$ is the leakage current through the cable parasitic capacitance.

If the measured and the simulated currents are the same,

$$I_{cha}(t) - I^*_{cha}(t) = 0, \qquad (6)$$

$$I_{chb}(t) - I^*_{chb}(t) = 0, \qquad (7)$$

then the bit exchange is secure. If the currents are different, an attack may take place. If the difference is greater than a pre-agreed threshold value, Alice and Bob discard the bit.

The simulated comparison results at Alice's side are shown in Fig. 7. The solid line indicates a current injection attack and the $I_{cha}(t) - I^*_{cha}(t)$ difference is well visible. Alice and Bob can recognize the attack virtually immediately. The dashed line shows the secure situation with $I_{cha}(t) = I^*_{cha}(t)$.

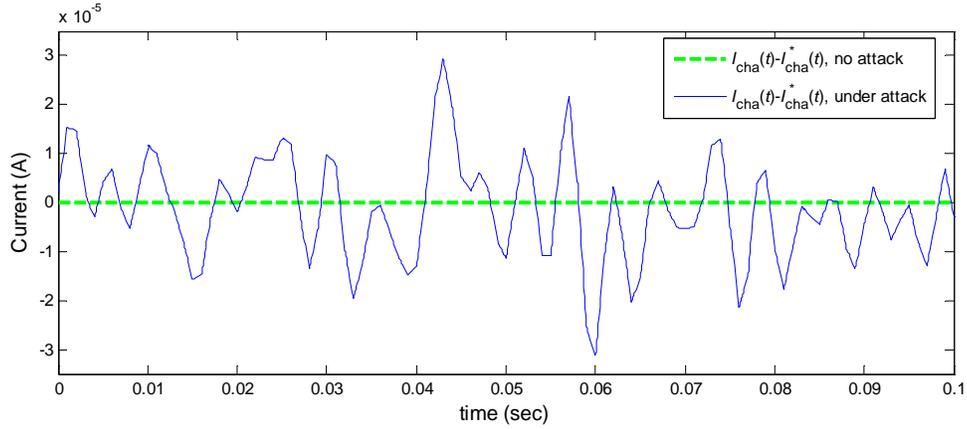

Fig. 7. Demonstration of the efficiency of the defense protocol with the practical cable over the bit exchange period. Alice and Bob can recognize the attack virtually immediately. The cable length is 1000 m.

## 4.2. Privacy Amplification

Privacy amplification is a well-known method that can be used to reduce any type of information leak [43]. The KLJN system can reach extraordinarily low bit error probability [38-40] thus privacy amplification (which is basically an error enhancer) can be efficiently be used. The simplest technique is the XOR-ing of the subsequent pairs of the key bits, that is, generating a new key which is cleaner and have half of the length of the original key. We simulated the effect of this technique at the most effective attack scenario, see Table 1. The simulation results showed that by XOR-ing once, Eve's success probability was reduced from 0.613 to 0.530, which was further reduced to 0.502 by XOR-ing the second time. The resulting key length became one quarter of its original length with significantly higher security.

## 5. Conclusions

In this paper, we validated the current injection attack against both the ideal and the practical KLJN system by utilizing LTSPICE. We have shown that the current and voltage comparison method, combined by in-site cable simulations, can efficiently detect and eliminate the attack.

## Author Contributions

Theory: Hsien-Pu Chen and Laszlo B. Kish; Math analysis: Hsien-Pu Chen and Laszlo B. Kish; Simulations: Hsien-Pu Chen; Interpretation: Hsien-Pu Chen and Laszlo B. Kish; Writing: Hsien-Pu Chen, Muneer Mohammad and Laszlo B. Kish. All authors have read and approved the final manuscript.